\documentstyle[12pt]{article} 

\oddsidemargin =  \evensidemargin
\newcommand{\eg}{{\it e.g.}}
\newcommand{\ie}{{\it i.e.}} 

\newcommand{\s}{\sigma}
\newcommand{\ap}{\alpha^{\prime}} 
\newcommand{\be}{\begin{equation}}
\newcommand{\ee}{\end{equation}} 
\newcommand{\ba}{\begin{eqnarray}}
\newcommand{\ea}{\end{eqnarray}}

\begin{document} 
 
\def\pct#1{(see Fig. #1.)} 



\begin{titlepage} \hbox{\hskip 12cm ROM2F-97/52  \hfil} 

\begin{center}  

{\Large  \bf  OPEN \ STRINGS \ AND \ DUALITIES} \vspace{0.6cm}

{\Large  Massimo BIANCHI } \vspace{0.4cm}

{\sl Dipartimento di Fisica, \ \ Universit{\`a} di Roma \ ``Tor Vergata'' \\
I.N.F.N.\ - \ Sezione di Roma \ ``Tor Vergata'', \ \ Via della Ricerca 
Scientifica, 1 \\ 00133 \ Roma, \ \ ITALY} \vspace{0.6cm}
 
ABSTRACT

\end{center}

{In the fruitful interplay between gauge fields and strings and in many 
conjectured M-theory dualities, open strings play a prominent role. We
review the construction of open-string descendants ({\it un-orientifolds}) of
closed-string theories admitting a generalized orientation reversal involution.
We then specialize the construction to some classes of non-supersymmetric models
in $D=10$ that have been recently considered in the context of duality without
supersymmetry. We also discuss the propagation of open and
unoriented strings on the NS pentabrane (N5-brane). This
background is a prototype of the configurations of branes and
orientifold planes that represent a powerful alternative to the {\it geometric
engineering} of Supersymmetric Yang-Mills Theories.  The resulting description
of D-branes in non-trivial backgrounds looks very different from the one naively
expected. In particular the very distinction between different Dp-branes 
becomes ambiguous in the presence of strong curvature effects.}

\vspace{0.6cm} 

\end{titlepage} 


\section{Introduction}

\subsection{In The Web of String Dualities}

  At the perturbative level, there are two rather distinct classes of 
superstring theories: those with only closed oriented strings (type IIA, 
type IIB, $E(8)\times E(8)$ and $SO(32)$ heterotic) and those with open and
closed unoriented strings (type I) \cite{gsw}.  In the past, models with open
strings have been studied to a lesser extent than models with only 
closed oriented strings. Though the initial proposal \cite{as} of identifying
open-string theories as {\it parameter-space orbifolds} of left-right symmetric
theories of closed oriented strings has been brought to a fully consistent
systematization long time ago \cite{bssys}, for almost ten years
{\it phenomenological} considerations  have oriented the interest of the string
community towards perturbative vacua of the $E(8)\times E(8)$ heterotic string
preserving ${\cal N}=1$ supersymmetry in $D=4$ \cite{gsw}. After the work of
Seiberg and Witten on ${\cal N}=2$ supersymmetric Yang-Mills theories \cite{sw},
these motivations seem to become much less compelling and a rather appealing
scenario is taking shape according to which different string theories should be
regarded as dual manifestations of a more fundamental M/F-theory \cite{mt}.
 
The most fashionable possibility is a theory of membranes (2-branes) and
penta-branes (5-branes), known as M-theory, that almost by definition has 11D
supergravity as its low-energy limit \cite{mt}.  This theory has no analogue
of the dilaton, whose (undetermined) vacuum expectation value plays the role of
the string coupling constant, and does not allow a perturbative expansion prior
to compactification. Upon dimensional reduction to  $D=10$ M-theory gives the
type IIA superstring \cite{witdyn}. Upon compactification on a segment it is
conjectured to give the $E(8)\times E(8)$ heterotic string \cite{horwit} on the
basis of symmetry between the two fixed ``points", surviving supersymmetry,
anomaly cancellation {\it and} a consistent string interpretation. Relaxing the
latter condition a theory such as ${\cal N}=(1,0)$ supergravity coupled
to $U(1)^{248}\times U(1)^{248}$ vector supermultiplets \cite{gsw} could not be
excluded.

In the intricate web of conjectured dualities a prominent role is played by the 
type II solitons carrying Ramond-Ramond (R-R) charges \cite{cjp}. These have 
been
formerly identified with charged black p-brane solutions \cite{dkl} of the 
string
equations of motion to lowest order in  $\ap$, the inverse string tension.  A
microscopic  description \cite{cjp} in terms of hyperplanes where open strings
can terminate with Dirichlet boundary conditions (D-branes) has opened the way
to remarkable  progress not only in the (string) duality realm but also in the
context of black-hole thermodynamics \cite{jm}. 

Much in the same way as the two type II superstrings are related by a 
non-geometrical $Z_2$-orbifold procedure through the action of $(-)^{F_L}$ and
similarly the two heterotic strings through  $(-)^{F_{16}}$, the general
construction of perturbative open-superstring vacuum configurations consists in
a non-standard $Z_2$-orbifold procedure that amounts to {\it gauging} the action
of the world-sheet parity operator $\Omega$ \cite{as}\footnote{Many authors
refer to these as {\it world-sheet orbifolds} \cite{hor} or as 
{\it orientifolds}
\cite{cjp} though the name {\it un-orientifolds} would be more suitable since 
the
resulting string is unoriented and carries no conserved charge.}.  From the
target-space viewpoint the $\Omega$-projection corresponds to the introduction
of an orientifold 9-plane (O9-plane) that roughly speaking induces a doubling of
the spacetime points.

Open strings and D-branes play a crucial role in many recently conjectured
string  dualities \cite{mbik}.  The $SO(32)$ type I superstring may be 
considered
as describing the excitations of a BPS configuration of 32 type IIB D9-branes
needed to neutralize the R-R charge of an  O9-plane. Similarly, the excitations
of the type I D-string (D1-brane) exactly coincide with  the (light-cone)
degrees of freedom of the $SO(32)$ heterotic string \cite{pw}. This observation
has strengthen the conjectured heterotic / type I strong-weak coupling duality
\cite{witdyn,dht}. It is not the purpose of the present talk to review type I
vacuum configurations in various  dimensions and their heterotic duals that
are analyzed in some detail in the talk of Carlo Angelantonj \cite{caik}. We
simply remind the reader that many non-perturbative features of heterotic vacua
allow for a quantitative description in the type I setting where open-string
excitations of D-branes are included in the perturbative spectrum. In this
respect, though all p-branes are belevied to stand on an equal footing in the
final non-perturbative formulation, open strings are representatives of a sort
of {\it open aristocracy} \cite{mbsig} since they play a privildged role as far
as perturbative description and explicit counting of BPS states are concerned.
  
The bridge between neutral {\it vacuum} configurations and charged
configurations has been provided by the study of D-brane probes \cite{dl}.
While the low-energy dynamics of a configuration of $N$ parallel Dp-branes is
governed by the dimensional reduction of ${\cal N}=(1,0)$
Supersymmetric Yang-Mills (SYM) theory with gauge group $U(N)$ from $D=10$ to
$D=p+1$ \cite{witdb} for other D-brane probes the dynamics is governed by
lower-SYM theories possibly coupled to matter supermultiplets \cite{dl}. Notice
that the Matrix Theory idea was born exactly in this context \cite{bfss}.  The
non-polynomial Dirac-Born-Infeld action \cite{cjp} has so far proven to be of
very limited usage so far due to the difficulty intrinsic to a non-abelian
generalization. It is remarkable that full Lorentz invariance is regained in the
continuum limit $N \rightarrow \infty$ of M(atrix)-theory without any need of
reference to the DBI formulation.

\subsection{The Geometric Origin of Field-Theory dualities}

Charged D-brane configurations and their open-string excitations have proven
useful at least in two respects. On the one hand they allow for a consistent
description of BPS configurations continuously  connected with charged extremal
black-holes that have allowed for a microscopic derivation of the
Beckenstein-Hawking entropy formula \cite{jm}. On the other hand they have been
used to explore the geometric origin of dualities in some SYM theories.

The  prototype of such configurations has been introduced by Hanany and
Witten \cite{hanwit}. It consists of a set of parallel D3-branes suspended
between two parallel N5-branes \cite{chs} in such a way that the 4D world-volume
of the D3-branes has 3 non-compact dimensions in common with the 6D world-volume
of the N5-branes. This configurations break 1/4 of the original supersymmetries,
\ie~preserve ${\cal N}=4$ supersymmetry in $D=3$. That D3-branes can end on
N5-branes is U-dual to the fact that D3-branes, that are $SL(2,Z)$ singlets, can
end on D5-branes, that transform into N5-branes under $SL(2,Z)$. A gas of
transversal D3-branes can be added that does not break any further
supersymmetry. Some remarkable properties of ${\cal N}=4$ SYM in $D=3$ can be
derived as a consequence of allowed movements of these configurations and of a
simple rule of {\it anomalous  creation} of D-branes \cite{acdb}. Performing
T-duality one can pass to configurations of N5-branes, D4-branes and D6-branes
that preserve ${\cal N}=2$ supersymmetry in  $D=4$ \cite{ejs}. Once again
allowed movements and {\it anomalous creation} of D-branes give rise to a
remarkable understanding of certain dualities in  ${\cal N}=2$ SYM theories. By
rotating one of the two N5-branes \cite{bar} one can even pass continuously from
a configuaration with ${\cal N}=2$ supersymmetry to a configuration with 
${\cal N}=1$ supersymmetry \cite{egk}. On the (3+1) non-compact directions of
the $N_4$ D4-branes the effective low-energy dynamics is governed by $U(N_4)$
SQCD with $N_6$ flavours. In practice the D6-branes are much heavier and their
dynamics is so slow that the $U(N_6)$ Chan-Paton (CP) group plays the role of a
global symmetry. Separating the D6-branes has a field theory analogue in the
introduction of superpotential mass-terms. Separating the D4-branes correspond
to Higgsing, \ie~giving vevs to the squarks. A similar correspondence may be
established for the ${\cal N}=2$ configurations. Seiberg's duality \cite{aps}
follows from transporting the D6-branes past one of the two N5-branes. In the
movement a certain number of D4-branes is created. Adding O-planes one can
analyze orthogonal and symplectic CP groups \cite{ejs}.  More recently the
dynamics of many configurations of branes and planes has been shown to be
equivalent to the world-volume dynamics of a BPS configuration of a single
M5-brane of complicated topology \cite{witmt}. In this M-theory approach the
hyperelliptic curves that appear in the determination of the Wilsonian action
for ${\cal N}=2$ SYM theories are simply the surfaces around which the M5-brane
world-volume is wrapped in order to give an effective 4D dynamics.

So far the dynamics of D-branes and O-planes has been studied mostly from a 
macroscopic point of view. In the second part of the talk we will address the
issue of a microscopic description of the excitations of D-branes
and  O-planes in curved backgrounds. For the sake of a quantitative analysis we
will restrict our attention to  much simpler configurations such as the
background of $k$ coincident N5-branes \cite{ejs,fgp,penta}. More interesting
configurations represent a challenge for future work in the field.

Some of the consistency requirements that are present in type I vacuum
configurations, such as tadpole cancellation \cite{cp,ps,bssys},
may be relaxed when the R-R charge can leak out at infinity. The non-compact 4D
background transverse to $k$ coincident N5-branes admits an exact ${\cal
N}=(4,4)$ superconformal field theory (SCFT) description in terms of an $SU(2)$
WZNW model at level $k$ and a Feigin-Fuchs (FF) boson with background charge
$Q=\sqrt{2/(k+2)}$ \cite{chs}. Moreover, as a consequence of the linear dilaton
background in the throat region accessible to CFT techniques, all closed-string
states become massive for any finite $k$ and the very definition of a tadpole
for an off-shell state is questionable in perturbative string theory
\cite{bsbos}. Nevertheless the factorization properties of the {\it tube}
amplitudes and the relation between direct (open-string) channel and transverse
(closed-string) channel are restrictive enough to fix the parametrization of the
open-string spectrum almost completely. Not counting T-dualities along possible
compact world-volume directions of the N5-branes, there are two different parent
type IIA configurations that admit the introduction of D-branes and O-planes. We
will discuss only the simplest configuration and refer the interested reader to
\cite{penta} for the other configuration. The former however already shows the
main features of the problem. The distinction between different kinds of
D-brane is ambiguous if not impossible for any finite $k$ due to the strong
curvature of the background. From the underlying CFT viewpoint the ambiguity
manifests itself in the non-abelian structure of the fusion algebra. Another
amusing result follows the introduction of a magnetic field in the above
non-trivial background. It induces a twisting of the $SU(2)$ current algebra 
in the open-string sector with the consequent breaking of target-space
supersymmetry \cite{acny}.

For the sake of comparison, in the first part of the talk, we will present
some $D=10$ models where unphysical tadpole cancellation go hand by hand  with
anomaly cancellation \cite{cp,bssys}. On the contrary, though tree-level vacuum
stability tends to favour backgrounds with no dilaton tadpole
\cite{bsbos,bssys}, tadpoles for physical massless states, such as the dilaton,
may be disposed of in principle via the Fischler-Susskind (FS) mechanism
\cite{fs} and have no relation whatsoever to anomalies. 

The plan of the talk is as follows. In the next section we describe the general
strategy for deriving open-string descendants of Rational Conformal Field
Theories (RCFT) and discuss some non-supersymmetric models in $D=10$. Their
relevance to some recently proposed dualities that involve non-supersymmetric
interpolating string models \cite{bd} will be streamlined. We then pass to
charged configurations and apply the general strategy and some useful results 
for
open-descendants of $SU(2)$ WZW models \cite{wzw} to describe the propagation of
open and unoriented strings on a class of BPS configurations of N5-branes,
D-branes and O-planes. Finally we present lines for future investigation.

\section{Open-string Descendants of RCFT}
 
The introduction of D-branes and O-planes in a type II background requires the
identification of the corresponding boundary (B) and crosscap (C) states.
Consistency conditions for boundary states in curved backgrounds have been
recently discussed in \cite{oog}. The procedure in fact may be reversed. 
Starting
from a consistent type II background and more generally from a consistent CFT
that admit an automorphism $\Omega$ exchanging left and right movers one can
deduce the open-string descendant and then extract the information on the
admissible B and C states from the resulting  brane configurations. 
We briefly describe the general procedure\footnote{For a more detailed analysis
see \eg~ \cite{gpik} or the original literature \cite{bssys,bpstor,wzw} on
the subject.} and then specialize to some non-supersymmetric models in $D=10$.

\subsection{General Strategy}

The starting point of the construction is a left-right symmetric modular
invariant torus partition function \cite{as,bssys}.  
The characters of the chiral
algebra, that extends the (super)Virasoro algebra, 
\be 
\chi_h(q) = Tr_{_{{\cal H}_h}} q^{L_o-{c\over 24}} 
\label{spectrum}
\ee
encode the (chiral) closed-string spectrum of the RCFT and enter the torus
amplitude in a modular invariant way
\be  
{\cal T} = \sum N_{h\bar h} \chi_h \bar \chi_{\bar h}  \qquad .
\ee 
The $\Omega$-projection introduces O-planes that are accounted for by the
Klein-bottle amplitude    
\be 
{\cal K} = {1\over 2} \sum N_{hh} \s_h \chi_h \qquad . 
\ee 
The signs $\s_h$ are restricted by the crosscap
constraint \cite{fpss} that \eg~ requires $\s_i\s_j=\s_k$ if $[i]\times
[j]= N_{ij}^k [k]$. The most general parametrization of the open-string spectrum
involves the annulus partition function: 
\be 
{\cal A} = {1\over 2} \sum A_{ab}^h n^a n^b \chi_h  
\ee 
and the M\"obius strip projection 
\be 
{\cal M} = {1\over 2} \sum M_{aa}^h n^a \widehat\chi_h  
\ee 
where $\widehat\chi_h$ form a proper basis of {\it hatted} characters
\cite{bssys}  
\be
\hat\chi_h (i\tau_2 + 1/2) = exp(-i\pi (h-c/24)) \chi_h(i\tau_2 + 1/2) \qquad .
\ee
Though the CP indices $a$ and the
character indices $h$ vary in general over different sets, for
the charge-conjugation modular invariant one is allowed to take them in the same
range and let $A_{ij}^k=N_{ij}^k$ or an automorphism thereof
\cite{bssys}. Sewing of surfaces with holes and crosscaps implies some
consistency conditions on the above parametrization. Most notably
$A_i^a{}_c A_j^c{}_b=N_{ij}{}^kA_k^a{}_b$ \cite{fpss}. After switching to the
transverse closed-string channel via a modular S-transformation, ${\cal K}$
yields the crosscap-to-crosscap amplitude 
\be 
\widetilde{\cal K} = \sum |\Gamma_h|^2 \chi_h  
\ee 
and ${\cal A}$ yields the boundary-to-boundary amplitude  
\be 
\widetilde{\cal A} = \sum (B^h)^2\chi_h  = \sum (B_a^h n^a)^2 \chi_h \qquad .
\ee 
Given these two amplitudes, a consistency check arises from the
bondary-to-crosscap amplitude that must be of the form 
\be 
\widetilde{\cal M} = \sum \Gamma_h (B_a^h n^a) \widehat\chi_h 
\ee
where the $\Gamma$'s hide some sign ambiguity.
The modular transformation between loop and tree channel of the M\"obius
strip is induced by  $P = T^{1/2} S T^2 S T^{1/2}$ that acts on hatted
characters and satisfies $P^2 = C$ \cite{bssys}. The boundary reflection
coefficients $B_a^k$ satisfy polynomial equations \cite{fpss} that in general
constitute an overcomplete set that determines the allowed boundary states. In
its naivest form the equations read 
\be 
B^{(a)}_i B^{(a)}_j = N_{ij}^k B^{(a)}_k 
\ee 
where the index $a$ labels the independent solutions \ie~ the independent
CP multiplicities. One thus has two alternatives. Either one starts from the
direct cannel and determines the signs $\s_h$ in ${\cal K}$ and the coefficients
$A_{ab}^k$ in ${\cal A}$ imposing consistency of the transverse channel, or
solves the polynomial equations for the reflection coefficients $B$'s,
parametrizing the transverse channel amplitudes, and then requires that the
direct channel be compatible with the CP multiplicities $n$'s being integers.

In order for un-orientifolds of RCFT's to describe consistent open-string
vacuum configurations two more requirements are to be met. The first is the
correct relation between spin and statistics \cite{bsrmg}. The second is the
cancellation of the tadpoles of unphysical massless states \cite{cp,bssys}.
In the following we will show that while in supersymmetric theories
tadpoles of physical, \eg~ the dilaton, and unphysical closed-string massless
states are not independent, in non-supersymmetric theories the two are often
independent and anomalies are related only to the unphysical massless tadpoles.

\subsection{Non-supersymmetric Models in $D=10$}

The simplest rational closed-string theory that admits an
open-string descendant is the type IIB theory in $D=10$. The result is the type 
I superstring with gauge group $SO(32)$ \cite{as}. In $D=10$ there are two more
left-right symmetric theories that admit open-string descendants \cite{bssys}.
The parent closed-string theories are tachyonic and are termed type 0A and type
0B \cite{bd}, in order to display the lack of supersymmetry. These theories are
non-geometrical orbifolds of the type IIA and type IIB superstrings with
respect to the $Z_2$-projection generated by $(-)^{F_L + F_R}$. Up to irrelevant
factors, the closed-string spectrum is encoded in   
\ba   
{\cal T}_A &=& |O|^2 + |V|^2 +S\bar{C} + C\bar{S} \\   
{\cal T}_B &=& |O|^2 +|V|^2 + |S|^2 + |C|^2 
\label{nonsusy} 
\ea 
where $\{ O,V,S,C\}$ are the characters of the $SO(8)$ transverse Lorentz 
current
algebra at level one \cite{gpik}. The conventional Klein-bottle projections 
\ba 
{\cal K}_A &=& O + V \\ 
{\cal K}_B &=& O + V - S - C    
\ea 
do not remove the closed-string tachyons from the spectrum. The massless
bosonic spectrum of the A-model contains the graviton and the dilaton in the
NS-NS sector together with a vector and a 3-form potential in the R-R sector. 
One may thus expect charged Dp-branes with $p=0,2,4,6$.  The massless bosonic
spectrum of the B-model contains the graviton and the dilaton in the NS-NS
sector together with two 2-form potentials in the R-R sector. One may thus
expect two independent sets of charged Dp-branes each with $p=1,5,9$.  From the
tree-channel Klein bottle amplitudes that reads  
\ba 
\widetilde{\cal K}_A &=& 32 \times (O + V) \\
\widetilde{\cal K}_B &=& 2 \times 32 \times V 
\ea
one deduces that the O9-planes neither carry R-R charge in the A-model nor in 
the
B-model. A similar analysis of the transverse annulus amplitudes reveals that no
R-R charge neutrality conditions emerge from unphysical tadpoles in the A-model,
consistently with the absence of D9-branes. On the contrary in the B-model there
are two unphysical D9-brane tadpoles to cancel. The resulting CP groups are
$SO(N)\times SO(M)$ and $SO(N)^2\times SO(M)^2$ respectively. The doubling of
the CP symmetry is clearly related to the doubling of D9-branes! In the A-model
the open-string spectrum includes tachyons in the adjoint or in the symmetric
tensor representation, depending on an unconstrained sign in the M\"obius
strip, and non-chiral spinors in the representation $({\bf N},{\bf M})$. In the
B-model the  open-string spectrum includes tachyons in the representations
$({\bf N},{\bf N},{\bf 1},{\bf 1})$ and $({\bf 1},{\bf 1},{\bf M},{\bf M})$,
left spinors in the representations $({\bf N},{\bf 1},{\bf M},{\bf 1})$ and
$({\bf 1},{\bf N},{\bf 1},{\bf M})$ as well as right spinors in the
representations $({\bf N},{\bf 1},{\bf 1},{\bf M})$ and $({\bf 1},{\bf N},{\bf
M},{\bf 1})$. The irreducible anomaly cancels thanks to the mirror-like
structure of the fermion representations. The reducible part does not factorize
and requires the contribution of both R-R antisymmetric tensors \cite{gss}.
Notice that for $N+M=32$ the dilaton tadpole cancels and the flat Minkowsi
vacuum is stable with respect to the lowest order (genus one half) quantum
corrections. We stress again that the dilaton tadpole has no relation whatsoever
to the anomaly and in principle it can be disposed of through the
Fischler-Susskind mechanism \cite{fs} since the dilaton is a physical state of
the projected unoriented closed-string spectrum.  

Keeping in mind that in the covariant extension of the operator content of
these models it is $V$ that plays the role of the identity in the fusion
algebra, one easily finds that there are other {\it unconventional} Klein-bottle
projections compatible with  the crosscap constraint \cite{fpss}:  
\ba 
{\cal K}_{A1} &=& V - O  \\ 
{\cal K}_{B1} &=& V + O + S + C \\ 
{\cal K}_{B2} &=& V - O - S + C \\ 
{\cal K}_{B3} &=& V - O + S - C  
\ea
The last two are equivalent under 10D parity. In the models A1 and B2 the
closed-string tachyon is projected out. In the R-R sector one has a vector and a
3-form in the A1-model, two scalar and a non-chiral (with no fixed duality
properties) 4-form in the B1-model and a scalar, a 2-form and a self-dual 4-form
in the B2-model. One thus expects the following charged Dp-branes. In the
A1-model $p=0,2,4,6$. In the B1-model two sets with $p=-1, 3, 7$. Finally in the
B2-model $p=-1,1,3,5,7,9$. Notice that up to uninteresting factors, the
transverse channel amplitudes read  \ba 
\widetilde{\cal K}_{A1} &=& - 32 \times (S+C) \\ 
\widetilde{\cal K}_{B1} &=& +2\times 32 \times O   \\
\widetilde{\cal K}_{B2} &=& -2\times 32 \times S  \\ 
\widetilde{\cal K}_{B3} &=& -2\times 32 \times C 
\ea 
The A1-model is manifestly
inconsistent with the direct channel, \ie~ with 10D Lorentz invariance and the
absence of D9-branes. The B2(B3)-model require D9-branes while in the B1-model
the O9-planes do not carry R-R charge. The open-string spectrum of the B1-model
displays a $U(N)\times U(M)$ CP symmetry and includes tachyons in the 
representations $({\bf N(N \pm 1)/2 + c.c., 1}) $ and $({\bf 1, M(M \mp
1)/2+c.c.})$, left-handed fermions in the  representation $({\bf N, M})+c.c.$
and right-handed fermions in the representation $({\bf N, M^*})+c.c.$. The
non-abelian part of the irreducible anomaly  cancels thanks to the mirror-like
structure of the fermion representations. The abelian part is made innocuous by
the analogue of the Dine-Seiberg-Witten (DSW) mechanism \cite{dsw} that allows
the abelian open-string vector to become massive by its coupling to the R-R
scalar. After the decoupling of the massive photon there is no left-over
reducible anomaly. The open-string spectrum of the B2-model displays a
$U(N)\times U(M)$ CP symmetry and includes tachyons in the representation
$({\bf N, M^*})+c.c.$ , left-handed fermions in the representations $({\bf N(N -
1)/2 + c.c., 1})$ and $({\bf 1, M(M + 1)/2+c.c.})$  and  right-handed
fermions in the representation $({\bf N, M})+c.c.$. Tadpole cancellation
require $N-M=32$. The anomaly is partly cancelled by the GSS mechanism and
partly by the DSW mechanism. Notice that there is no choice of $N$ and $M$ that
allows the cancellation of the dilaton tadpole. In principle it has to be
disposed of via the FS mechanism.

The interest in these models is twofold. On the one hand they neatly illustrate
the procedure and the relation betwen D-branes, CP charges and
tadpoles in flat spacetime. On the other hand they play a significant role in
some recently proposed string dualities without supersymmetry \cite{bd}. Most
dualities have so far relied heavily in the tight constraints supersymmetry
imposes on the spectra and interactions and one has the right to wonder if
string dualities are a property to be associated to the existence of
non-pointlike constituents or are better an unescapable property of the
low-energy supergravity, that however clearly {\it knows} about the presence of
branes. Though based on the continuous connection in $D=9$ between
non-supersymmetric and supersymmetric models, the dualities proposed in
\cite{bd} represent the first instances of a much wider context.  In particular
from the open-string perspective only a soliton interpolating between
$SO(16)\times SO(16)$ and $SO(32)$ seems to play a role in the duality pattern
so far proposed \cite{bd}. The other consistent open-string models described
above should eventually find a {\it raison d'\`etre} that hopefully might
provide a string interpretation for the ${\cal N}=(1,0)$ supergravities with
gauge groups $U(1)^{496}$ or $E(8)\times U(1)^{248}$ in much the same way as the
strong coupling limits of the type IIA and $E(8)\times E(8)$ heterotic strings
provide string interpretations for the ${\cal N}=1$ supergravity in $D=11$
\cite{witdyn,horwit}. 

Undoubtly the continuous connection between supersymmetric and 
non-supersymmetric  string
models may prove very useful both in the study of black hole thermodynamics,
where extremal or nearly extremal solutions have attracted so far
most of the attention \cite{jm}, and in the study of
non-supersymmetric YM theories, where the problem of color confinement and
chiral symmetry breaking are still poorly understood by analytic means.

\section{From Spheres to Pentabranes}

Having described the general strategy to derive open-string descendants of
left-right symmetric closed-string models based on RCFT and discussed some
unconventional open-string models in $D=10$ to highlight the relation with the
by-now standard D-brane technology \cite{cjp} and string dualities \cite{bd},
we would now like to turn to charged configurations of D-branes where the RR
charges may leak out at infinity and the very distinction between
different Dp-branes becomes rather fuzzy due to strong curvature effects.
Following \cite{penta} very closely, we will discuss open string propagation in
the background of $k$ coincident NS pentabranes (N5-branes). N5-branes
are string solitons with NS-NS magnetic charge and may be visualized as extended
objects with a $5+1$-dimensional worldvolume \cite{chs}. After setting
to zero all the R-R fields, the background of type II N5-branes is
completely determined by  
\ba 
ds^2 &=& \eta_{\mu\nu} dx^{\mu} dx^{\nu} + e^{-2\phi} (dr^2 + r^2 ds^2_3) \\ 
e^{-2\phi} &=&  e^{-2\phi_o}  (1 + {k\over r^2}) \\ 
H &=& dB = * de^{-2\phi} = - k d\Omega_3 \quad ,  
\label{pentabrane} 
\ea
where the indices $\mu,\nu = 0,1,2,3,4,5$ are tangent to the N5-brane, $ds^2_3$
and $d\Omega_3$ are the line and volume elements on $S^3$, respectively. The
geometry of the space transverse to the N5-brane is that of a semi-wormhole with
the size of the throat fixed by the axionic charge $k$ (number of coincident
N5-branes). In the throat region, $r\rightarrow 0$, where the dilaton diverges,
the N5-brane background (\ref{pentabrane}) admits an exact CFT description as 
the
tensor product of the $SU(2)$ WZNW model at level $k$ times a FF boson
$X_4$ with background charge $Q=\sqrt{2/k+2}$ \cite{chs}. Including the
fermionic partners $\{\psi^i, \psi^4\}$, the world-sheet theory enjoys an
extended  ${\cal N}=(4,4)$ superconformal symmetry which guarantees the absence
of both perturbative and non-perturbative corrections in $\alpha^{\prime}$
\cite{chs}.  The complete construction of the modular invariant
spectrum of closed-string excitations around the semi-wormhole background for
even values of $k$ has been worked out in \cite{afk}, where other classes of 4-d
backgrounds with exact ${\cal N}=(4,4)$ superconformal symmetry have been 
constructed.

Many other exact 4D backgrounds (generalized hyper-k\"ahlerian manifolds) and
their T-duals have been analyzed in \cite{kkl} in relation to non-compact
Calabi-Yau manifolds and axionic instantons. Stringy ALE and ALF instantons and
their behavior under Buscher duality \cite{bus} were thoroughfully analyzed in
\cite{ale}. More recently, string dualities in $D=6$ have been given support by
the observation that type IIA (B) with $k$ coincident N5-brane is equivalent
(Buscher T-dual) to type IIB (A) around an ALE space  $R^4/\Gamma_k$ \cite{ov}
at vanishing B-field \cite{asp}. The crucial observation is that the
(non)-chiral type IIB(A) admit N5-branes with
${\cal N}=(2,0)$, respectively ${\cal N}=(1,1)$, supersymmetry \cite{chs}. 
For the type IIB
N5-brane one expects an ${\cal N}=(1,1)$ vector multiplet whose four scalar 
components
are the collective coordinates for the translation of the N5-brane in the
transverse 4D space. This fits in with the conjectured $SL(2,Z)$ U-duality of 
the
type IIB superstring in $D=10$ which relates the N5-brane to the D5-brane,  the
world-volume degrees of freedom of the latter being open-string massless
excitations in an ${\cal N}=(1,1)$ vector multiplet. On the contrary, the type 
IIA
N5-brane requires an ${\cal N}=(2,0)$ tensor multiplet with 5 scalars, that are 
very
suggestive of an 11D interpretation in terms of M-theory. Indeed the type IIA
N5-branes are conjectured to arise from M5-branes by simple dimensional
reduction. On the contrary M5-branes wrapped around the eleventh dimension
should give rise to D4-branes. As a matter of fact, we will show that an
open descendant of the type IIA N5-brane can be consistently derived.
A previous analysis, that we have not been able to reconcile with the 
non-abelian
structure of the fusion algebra of the underlying SCFT at $k\neq 1$, has been
performed in \cite{fgp} and partly anticipated in \cite{ejs}. 

After an anomalous chiral rotation the world-sheet fermions $\Psi^i$
decouple from the $SU(2)$ currents and the level gets shifted $k \rightarrow
k-2$. The problem of un-orientifolding the N5-brane in an a priori flat
spacetime can be directly mapped to the already solved problem of
un-orientifolding the $SU(2)$ WZNW models \cite{wzw}. Indeed, the contributions
of the FF boson $X^4$, the flat bosonic coordinates $X^\mu$ and the fermions
$\{\psi^\mu, \psi^i, \psi^4 \}$ to the torus partition function is rather
trivial. Neglecting the discrete
representations that are finite in number and thus do not contribute to the
partition function but play the role of screening charges in the computation of
scattering amplitudes, the FF boson and each non-compact coordinate give the
standard contribution ${(\sqrt{\tau_2} |\eta|^2)}^{-1}$. 

For N5-branes in a priori flat spacetime the torus partition function 
factorizes 
\be
{\cal T} = (V -S) (\bar V  - \bar C) \sum_{ab} {\cal I}_{ab} \chi_a
\bar\chi_b \label{torus} 
\ee 
where $\{O,V,S,C\}$ have been introduced in eq. (\ref{nonsusy}) ${\cal I}_{ab}$
is one of the $A-D-E$ modular invariant combination of $SU(2)$ characters
\cite{ciz}. Since ${\cal I}_{ab}$ are all left-right symmetric one can always
perform an $\Omega$-projection. Notice that $\Omega$ combines world-sheet
left-right interchange with the $SU(2)$ involutions $g\rightarrow \pm g^{-1}$
and a flip of the chirality of the spinors $S_8 \rightarrow C_8$ \cite{ejs}.
This is consistent with the following observation. Since the volume of the
throat of the wormhole is quantized in units of $k$. One can use overall factors
of $k$ to trace the scaling of the open and unoriented amplitudes with the
volume and identify the configurations of branes and planes in the large volume
limit.  For finite $k$ the distinction between different Dp-branes becomes less
compelling due to strong curvature effects. Let us anticipate the results for
the two descendants of the $A$-series. We will see that for the parametrization
with real CP charges, keeping the lowest lying states with $a < \sqrt{k}$ the
amplitudes are independent of $k$ for large $k$ plus subleading terms. This we
interpret as an indication that the unorientifold introduces D6-branes and
O6-planes in this case. For complex CP charges, one finds amplitudes that up to
subleading terms behave as $k$ for large $k$. This we interpret as an indication
that the unorientifold introduces D8-branes and O8-planes in this case. Tadpoles
for the type IIA R-R 9-form and 7-form need not be cancelled because the RR
charge can flow to infinity and moreover all the states in the closed-string
spectrum in the throat region get a mass from the linear dilaton background.  

Since the only non-trivial contribution
to string propagation in the throat of the N5-brane is given by the
$SU(2)$ part, let us briefly recall for completeness some known facts about 
$SU(2)$ WZNW models and their open-string descendants. Open
string propagation on $S^3$ has been considered in connection to 2-d charged
black holes \cite{hor}. The problem was throroughly addressed along the lines of
\cite{bssys} and completely solved in \cite{wzw}.
The central charge of the Virasoro algebra is $c = 3k/(k+2)$ and the conformal
weights of the integrable unitary representations are   
\be 
h_j^{(k)} = {j(j+1)\over k+2} \label{wieght}  
\ee 
with isospin $j$ in the range $j=0,..k/2$. 
The generalized character formula is given by  
\ba
\chi_j^{(k)}(\tau,z,u) &=& Tr_{{\cal H}_j^{(k)}} 
q^{L_o-{c\over 24}} e^{2\pi i z J_o^{(3)}} = \\      
&=&  e^{2\pi i k u} { {q^{h_j^{(k)}-{c(k)\over 24}}
\sum_{n} q^{(k+2)n^2+(2j+1)n} sin[\pi z (2j+1+2 n(k+2))]} \over {sin (\pi z)
\prod_{n=1}^{\infty} (1-q^n) (1-e^{2\pi i z} q^n)
(1-e^{-2\pi iz}q^n)}}                         
\label{character} 
\ea  
For later purposes it is
convenient to label states and characters in terms of the
dimension of the corresponding heighest weight $SU(2)$ representations
$a=2j+1$.  The modular transformations in the above basis are representated
by the  matrices   
\be
S_{ab} \  = \  \sqrt{{2 \over k+2}}  \ sin \left({\pi a b  \over k+2}\right) \ ,
\label{smatrix} \ee 
and 
\be T_{ab} \  =  \ \delta_{a b} \  e^{i \pi \left( {a^2
\over 2(k+2)} - {1 \over 4}\right)} \quad . \label{tmatrix} \ee 
The charge
conjugation matrix is equal to the identity  $C=S^2 = (ST)^3 = 1$.  The modular
transformation between loop and tree channel of the M\"obius strip is induced
by  $P = T^{1/2} S T^2 S T^{1/2}$, for the $SU(2)$ WZNW model $P$ is represented
by \be  P_{ab} \  =  \ {2 \over \sqrt{k+2}} \  sin \left( {\pi a b  \over
2(k+2)}\right)
 \ (E_k  E_{a+b} + O_k O_{a+b}) \ , 
\label{pmatrix} 
\ee 
where $E_n$ and $O_n$
are projectors on even and odd $n$  respectively. The fusion rule coefficients
are given by  the Verlinde formula \cite{ev} \be 
N_{ab}{}^c \  =  \ \sum_{d=1}^{k+1} \ {S_{ad} S_{bd}
S^{\dagger}_{cd} \over S_{1d}}\quad , 
\label{nfusion} 
\ee 
$N_{ab}{}^c$ are non-zero, in fact equal to one, only for 
$|a-b|+1\le c \le min(k+1,a+b-1)$.
It turns out to be
convenient to introduce also the integer (!) coefficients  \be {Y_{ab}}^c  \ = 
\ \sum_{d=1}^{k+1}  \ {S_{ad} P_{bd} P^{\dagger}_{cd} \over S_{1d}} \quad .
\label{yfusion} \ee 

For simplicity of description let us restrict our attention to the open
descendants of the diagonal modular invariant, associated to $A_{k}$, that is
available at any level $k$.  The torus partition function reads 
\be 
{\cal T} \ = 
\ \sum_{a=1}^{k+1} | \chi_a |^2 \quad . \label{aseries} \ee  Corresponding to 
the
two geometrical involution on the $SU(2)$ group manifold, \ie~ $g\rightarrow
\pm g^{-1}$, there are two different $\Omega$-
projections  (Klein bottle amplitudes) of the parent torus partition function.
The two involutions have the same action on integer isospins, that trivialize
the center of $SU(2)$.  On the half-integer isospins, the former (-) involution
corresponds to keeping the antisymmetric part (\eg~ the singlets) of the 
diagonal
$SU(2)$ subgroup of the parent $SU(2)_L\times SU(2)_R$ symmetry. The latter (+)
keeps the symmetric states and removes \eg~ the singlets. We shall label the two
choices by an index $R$ and $C$ in order to streamline their relation to real
(orthogonal or symplectic) and complex (unitary) CP charge assignments. The
diagonal $A_{k}$ models allow for the introduction of  $k+1$ CP charges $n^i$ or
equivalently $k+1$ indipendent boundary states $B^{(i)}$ that are in one to one
correspondence with the integrable $SU(2)$ representations \cite{wzw}.

For the $A$-series  with real CP charges the Klein-bottle (${\cal K}_R$), 
annulus
(${\cal A}_R$) and M\"obius strip (${\cal M}_R$) direct (``loop") channel
amplitudes read
\ba
  {\cal K}_R  \ &=& \ \sum_{a=1}^{k+1} \  {Y^a}_{11}  \chi_a \ = \
\sum_{a=1}^{k+1} (-1)^{a-1} \chi_a \quad , \\ {\cal A}_R \ &=&
\sum_{a,b,c=1}^{k+1} {N_{ab}}^c \chi_c n^a n^b\quad ,   \\ {\cal M}_R \ &=& \
\pm \sum_{a,b=1}^{k+1} {Y_{a1}}^{b} \hat \chi_b n^a  \ = \ \pm
\sum_{a,b=1}^{k+1} (-1)^{a-1} (-1)^{b-1 \over 2}  {N_{aa}}^b \hat \chi_b n^a
\quad . \label{areald} \ea 
A modular transformation yields the transverse
(``tree") channnel amplitudes that are consistent with their interpretation as
closed-string amplitudes between boundary and/or crosscap states.

For the $A$-series with complex CP charges the various contributions to  the
direct channel partition function read 
\ba 
{\cal K}_C \ &=&   \ \sum_{a=1}^{k+1}
\  {Y^a}_{k+1,k+1}  \chi_a  \ = \ \sum_{a=1}^{k+1} \chi_a \quad ,  \\ 
{\cal A}_C \ &=&  \ \sum_{a,b,d=1}^{k+1} {N_{ab}}^d\chi_{k+2-d}  n^a n^b\quad , 
\\ 
{\cal M}_C \ &=&  \ \pm \sum_{a,b=1}^{k+1} {Y_{a,k+1}}^{b} \hat \chi_b n^a  \
= \ \pm \sum_{a,b=1}^{k+1}  {N_{aa}}^b \hat \chi_{k+2-b} n^a \quad . 
\label{acomplexd} 
\ea 
The transverse channel amplitudes are manifestly compatible with the required
factorization properties.  Notice that positivity of the transverse channel
requires the numerical identifications $n_{k+2-a} = \bar n_a = n_a$.

Since all closed-string states are massive it is not needed and in fact
plagued by ambiguities imposing the vanishing of any tadpole. It is 
questionable even to interpret the CP symmetry as a gauge symmetry since
for any finite $k$ the open-string vector multiplets are massive too.
Indeed, the spectrum is non-chiral and the R-R charge, if any, can leak
out at infinity with no bearing to anomaly consistency conditions. It is worth
stressing once again that the proper identification of the various Dp-branes
and Op-planes involved is possible only in the large $k$ limit, where the
regulating mass goes to zero too. This completes our discussion of the simplest
configuration of N5-branes, D-branes and O-planes. The open-descendants of the
other configuration alluded to in the introduction
has been analyzed in \cite{penta} where it was interpreted as arising from
placing the N5-branes at D-type orbifold singularities. For lack of space and
for the unavoidable technicalities involved in the discussion of this case we
refer the reader to \cite{penta}. For completeness we simply mention that the 
resulting open-string spectrum allows the presence of both massless and
tachyonic states. Choosing the CP factors in a proper way one can get rid
of the latter while keeping the former that seem to play the role of collective
coordinates of the bound state of D-branes and O-planes in the highly curved
background.    

\subsection{Adding a Magnetic Field}

As in toroidal and orbifold compactifications of open strings
\cite{bpstor,caik}, in the N5-brane background the
introduction of a constant abelian magnetic field \cite{acny} can be fully taken
into account. In the case at hand it corresponds to the insertion on the 
boundary
of the operator  \be  {\cal B}^i = J^i + {i\over 2} \epsilon^i{}_{jk} \psi^j
\psi^k 
\label{magnet} 
\ee 
This
boundary deformation of the rational CFT is integrable and one can express the
open-string spectrum in terms of the characters (\ref{character}) with $z$
related to the magnetic field ${\cal B}$ and the $U(1)$ charges of the
open-string state through 
\be 
z = {1\over \pi} ( arctg(q_1 {\cal B}) + arctg(q_2 {\cal B}) ) 
\label{shift}
\ee 
By $SU(2)$ symmetry one can always choose a ${\cal B}$ pointing along the
third direction. From the modular S-transformation 
\be
\chi_a^{(k)}(-{1\over\tau},-{z\over\tau},u-{z^2\over 2\tau}) = \sum_b S_{ab}
\chi_b^{(k)}(\tau,z,u) 
\ee 
one immediately deduces the Casimir energy and the
shift of the modes of the currents 
$J^{(\pm )}_N \rightarrow J^{(\pm )}_{N\pm z}$. Notice that, since the modes of
$J^{(3)}$ are uneffected, the current algebra is preserved 
\ba
 \left[ J^{(+)}_{n+z},J^{(-)}_{m-z} \right] &=& 2 J^{(3)}_{n+m} + k \delta_{n+m}
\\
\left[ J^{(3)}_{n},J^{(\pm )}_{m\pm z} \right] &=& \pm J^{(\pm )}_{n+m\pm z}
\\
 \left[ J^{(3)}_{n},J^{(3)}_{m} \right] &=& {k\over 2} \delta_{n+m}
\label{twistca}
\ea 
Indeed the introduction of the magnetic field simply amounts to a modulation of
the boundary reflection coefficients. By world-sheet supersymmetry
considerations an opposite shift of the modes is suffered by the  fermions.
However the total ${\cal N}=1$ supercurrent, \ie~ the one which couples to the
worldsheet gravitino,  
\be 
G = J^i \psi_i + i \partial X^4 \Psi_4 + {i\over 3!} \epsilon^{ijk} \Psi_i\Psi_j
\Psi_k + Q \partial \Psi_4  
\ee 
seems to forbids a twist of $\Psi_4$ (and
similarly of $X^4$) due to the presence of the background charge. The twisting
of only two currents and two fermions leads to an explicit breaking of the
spacetime supersymmetry. Since the curvature of the spin connection with torsion
is self-dual one may ask if there is any possibility of adding a self-dual
field-strength, \ie~ an instanton-like gauge field such as to make the 
background
supersymmetric. A possibility of this kind is reminiscent of the standard
embedding in the heterotic version of the N5-brane \cite{chs,ale}. This issue
claearly deserves further study. It may also prove interesting to explore its
generalization to compact curved background such as orbifolds, Gepner
models and fermionic models \cite{caik}. The final goal would be to address the
issue of consistency of magnetized D-branes inside Calabi-Yau (CY) manifolds
\cite{oog}.

\section{Final Comments}
 
Real progress in understanding  the dynamics of D-branes and O-planes 
in curved background 
may well benefit from exactly solvable CFT such as those described in
\cite{caik} or the string solitons described above.  D-brane instanton
corrections to superstring effective lagrangians clearly require a
precise understanding of the weighting and counting of D-branes wrapped around
non-trivial cycles in CY manifolds \cite{bbs}. Explicit computations in this
direction seem still rather difficult to perform. 

Recently, D-branes and their open-string excitations have
proven to be useful tools not only in the counting of microscopic degrees of
freedom of supersymmetric black-holes \cite{jm} but also in the {\it geometric
engeneering} of SYM theories \cite{witdb,dl,ejs,egk}. In
this context the dynamics of D-branes \cite{hanwit} and orientifold planes
\cite{ejs} in the background of symmetric penta-branes \cite{chs} leads to 
the anomalous creation of strings and branes \cite{acdb} and provides a
geometrical interpretation of some known dualities in SYM theories
\cite{aps}. A unifying picture emerges from \cite{witmt}, where  the relevant
configurations of N5-branes and D-branes are interpreted as a single M5-brane
wrapped around a Riemann surface.

Moreover, the very consistency of M-theory \cite{mt} and in particular of its
Matrix-Theory interpretation \cite{bfss} requires analyzing compactifications on
curved backgrounds that break some or all the supersymmetries. In this
respect the interpolating solitons studied in \cite{bd} provide a first step
towards a complete understanding of the consistent non-supersymmetric strings
\cite{gsw,bssys} that should play a role in the final non-perturbative scenario.
The lesson that we are learning from the fruitful interplay between
gauge fields, strings and branes is that any not manifestly
inconsistent theory, such as the resurrected 11D supergravity or the still
unexplored 10D supergravity with gauge group $U(1)^{496}$, deserves proper
attention and is not to be prematurely cut by means of {\it Ockam's razor}
\cite{razor}.

\section{Acknowledgements} 

The second part of this talk is based on work
done with Yassen Stanev that I would like to thank for an
enjoyable collaboration. I would like to acknowledge useful conversations with
C. Angelantonj, E. Kiritsis, G. Pradisi, S.-J. Rey, and A. Sagnotti and a
fruitful discussion concerning Ockam's razor with G. Preparata. I would also
like to express my deep gratitude to the local organizers of the {\it V Korean -
Italian Meeting on Relativistic Astrophysics} and particularly to Hyung-Won Lee
for offering their kind hospitality and for creating a stimulating enviroment.
Financial support has been provided by the KOSEF-CNR bilateral agreement.

\end{document}